\documentclass[12pt,twoside]{article}
\usepackage{emlines2}
\usepackage{amssymb}
\usepackage{wrapfig}
\usepackage{latexsym}
\usepackage{cmp2e}
\title[Scaling exponents of star polymers]%
{Field-theoretical renormalization group analysis for
the scaling exponents of star polymers}
\author[Ch. von Ferber, Yu.Holovatch]
{Ch. von Ferber\refaddr{label1}, Yu.Holovatch\refaddr{label2,label3}}
\addresses{
\addr{label1}
Theoretische Polymerphysik, Universit\"{a}t Freiburg,\\
D-79104 Freiburg, Germany
\addr{label2}
Institute for Condensed Matter Physics\\
of the National Academy of Sciences of Ukraine,\\
1 Svientsitskii Str., 79011 Lviv, Ukraine
\addr{label3}
Ivan Franko National University of Lviv,\\
79005 Lviv, Ukraine
}
\date{Received September 18, 2001}

\begin{document}
\setcounter{page}{117}
\maketitle

\begin{abstract}
We review recent results of the field theoretical renormalization
group analysis on the scaling properties of star polymers. We give a
brief account of how the numerical values of the exponents governing the
scaling of star polymers were obtained as well as provide some
examples of the phenomena governed by these exponents. In particular
we treat the interaction between star polymers in a good
solvent, the Brownian motion near absorbing polymers,
and diffusion-controlled reactions involving polymers.
\pacs 64.60.Ak, 61.41.+e, 64.60.Fr, 11.10.Gh
\end{abstract}


\section{Star exponents in polymer theory}\label{I}

It is well known that the asymptotic properties of a flexible
polymer chain in a good solvent are universal in the limit of
infinite chain length. Lowering the temperature in a system of a
polymer in a good solvent the so called $\Theta$ temperature may
be reached below which the polymer collapses. At the transition
temperature, the effective two point attractive and repulsive
interactions between  different monomers compensate each other
and as a result the polymer chain may be described by a random
walk (up to higher order corrections): the mean square distance
between the chain endpoint $\langle R^2 \rangle$ scales with the number of
monomers $N$ as $\langle R^2 \rangle\sim N$. Above the $\Theta$ temperature the
effective interaction between the monomers is repulsive resulting
in a swelling of the polymer coil which in the asymptotics again
obeys a universal scaling law:
\begin{equation} \label{1.1}
\langle R^2 \rangle\sim N^{2\nu}, \mbox{\hspace{4em} } N\rightarrow \infty
\end{equation}
with the universal exponent $\nu$ depending on the dimension of
space $d$ only.  The number of configurations $Z$ of a polymer
chain scales with  $N$ like
\begin{equation} \label{1.2}
Z\sim {\rm e}^{\mu N} N^{\gamma-1}, \mbox{ \hspace{4em}} N\rightarrow
\infty
\end{equation}
with a non-universal fugacity ${\rm e}^{\mu}$ and universal scaling
exponent $\gamma$.

In the early 70-ies, following the work of de Gennes
\cite{deGennes79}, the analogy between the asymptotic properties of
long polymer chains and the long-distance correlations of magnetic
systems in the vicinity of the 2nd order phase transition was
recognized and elaborated in detail. This mapping allows to define
the above exponents (\ref{1.1}), (\ref{1.2}) as $m\rightarrow 0$
limits of the correlation length critical exponent $\nu$ and the
magnetic susceptibility critical exponent $\gamma$ of the
$O(m)$--symmetric model. In this way, the powerful tools of the
field theoretical renormalization group approach to the
description of critical phenomena \cite{RGbooks} can be applied in
polymer theory (see e.g. \cite{polymerRG}). In particular, this
allows to calculate the exponents at $d=3$ with record accuracy
\cite{Guida98}:
\begin{equation}\label{1.3}
\nu(d=3)=0.5882\pm 0.0011 \mbox{\hspace{2em} and \hspace{2em}}
\gamma(d=3)=1.1596\pm0.0020.
\end{equation}
In dimension $d=2$, the exact values $\nu=3/4$ and $\gamma=43/32$
are known due to a mapping of two-dimensional (2D) polymers to the
2D Coulomb gas problem \cite{Nienhuis82}.

Under similar conditions (i.e. in good solvents above the $\Theta$
temperature and long polymer chains), star polymers \cite{note1}
also obey universal scaling laws. For the homogeneous star polymer,
the asymptotic properties are uniquely defined by the number of
its constituting chains and by the dimension of space
\cite{Duplantier89}. For the partition function (number of
configurations) $Z_f$ of a polymer star of $f$ chains, each
consisting of $N$ monomers, one finds:
\begin{equation} \label{1.4}
Z_f\sim {\rm e}^{\mu Nf} N^{\gamma_f-1}\sim(R/\ell)^{\eta_f-f\eta_2},
\mbox{ \hspace{4em}} N\rightarrow \infty.
\end{equation}
The second part shows the scaling in terms of the size $R\sim
N^\nu$ of the isolated coil of $N$ monomers on some microscopic
length $\ell$, omitting the fugacity factor. The exponents
$\gamma_f,\eta_f$, $f=1,2,3,\ldots$ constitute families of `star
exponents', which depend on the number of arms $f$ in a nontrivial
way.  The case of linear polymer chains is included in this family
with the exponent $\gamma=\gamma_1=\gamma_2$  defined in
(\ref{1.2}).

Apart from $\eta_f$ and $\gamma_f$ other sets of star exponents my
be defined that govern the scaling of different physical
properties of polymer stars. However, all these sets of exponents
can be expressed in terms of a given set by familiar scaling
relations \cite{Duplantier89}. Moreover, the scaling properties of
polymer networks of arbitrary but fixed topology are uniquely
defined by its constituting stars \cite{Duplantier89}, as long as
the statistical ensemble respects some conditions on the chain
length distribution \cite{Schaefer92}.  Thus, the knowledge of the
set of star exponents $\gamma_f$ or $\eta_f$ allows to obtain the
power laws corresponding to (\ref{1.4}) also for any polymer
network of arbitrary topology
\cite{Duplantier89,Schaefer92,FerHol97c}. Recently, the theory has
also been generalized to the case of star polymers and networks
that contain chains of different species, introducing additional
independent families of star exponents
\cite{FerHol97c,Duplantier99,FerHol01}.

As it is usual in the theory of critical phenomena, the space dimension
$d$ is crucial for the scaling laws (\ref{1.4}). At $d=2$, there
exist exact formulas expressing values of the scaling exponents of
the star polymers as functions of the number of arms
\cite{Duplantier86,Duplantier99,FerHol01}. In $d=3$ dimensions on
the other hand the problem does not allow for an exact solution
and different approximate schemes are used. The so-called cone
approximation takes into account the fact that for large $f$ each chain of
the star is restricted approximately to a cone of space angle
$\Omega_f=4\pi/f$. In this approximation one finds for large $f$
\cite{Ohno88}
\begin{equation}\label{1.5}
\gamma_f-1\sim -f^{3/2}.
\end{equation}
However, the most accurate values of critical exponents for $d=3$
are obtained by the field theoretical renormalization group
approach refined by a resummation of asymptotic series. Just in
this way high precision results for the exponents governing
different critical phenomena are derived \cite{RGbooks}, and this
is the way the scaling exponents (\ref{1.3}) for a polymer chain
in $d=3$ were obtained. In this review, we give a brief account on
how the numerical values of the star exponents were obtained for
$d=3$ by means of these techniques as well as we provide some
examples for phenomena governed by the star exponents.

The paper is arranged as follows. In the next section \ref{II} we
give a field-theoretical description of a star polymer in $d$
dimensions, the exponents governing scaling of a homogeneous star
polymer are evaluated numerically in section \ref{III}. The theory
is generalized to the case of a polymer star consisting of two
species of polymers (a copolymer star) in section \ref{IV}. The
following sections of the review are devoted to the numerical
description of different phenomena where star exponents come to
play: the interaction between polymer stars in a good solvent
(section \ref{VII}), the Brownian motion near absorbing polymers
(section \ref{V}), diffusion-controlled reactions in presence of
polymers (section \ref{VI}). We finally give some conclusions and
an outlook in section \ref{VIII}.


\section{Renormalization of polymer stars} \label{II}

To give a field-theoretical description of a star polymer let us
introduce an Edwards model of continuous polymer chain
\cite{polymerRG} generalized to describe a set of $f$ polymer
chains of varying composition, possibly tied together at their end
points. The configuration of one polymer is given by a path
$r^a(s)$ in $d$-dimensional space $I\!\!R^d$ parametrized by a
surface variable $0\leqslant s \leqslant S_a$. Let us allow for a symmetric
matrix of excluded volume interactions $u_{ab}$ between chains
$a,b = 1,\ldots,f$. The Hamiltonian $\cal H$ is then given by
\begin{equation}\label{2.1}
 \frac{1}{k_{\rm B}T} {\cal H}(r^a) =
\sum_{a=1}^{f}\int_0^{S_a}{\rm d}s \left(\frac{{\rm d}r^a(s)}{2{\rm
d}s}\right)^2 +
 \frac{1}{6}\sum_{a,b=1}^{f} u_{ab}\int{\rm
d}^d r \rho_a(r)\rho_b(r) ,
\end{equation}
with densities $\rho_a(r)=\int_0^{S_a}{\rm d}s
\delta^d(r-r^a(s))$. In this formalism the partition function is
calculated as a functional integral:
\begin{equation}\label{2.2}
 {\cal Z}_f\{S_a\} = \int{\cal D}[r^a(s)] \exp\left\{-
\frac{1}{k_{\rm B}T} {\cal H}(r^a)\right\} .
\end{equation}
Here, the symbol ${\cal D} [r_a(s)] $ includes normalization such
that $Z \{ S_a \} = 1$ for all $u_{ab}=0$. To make the exponential
of $\delta$-functions in (\ref{2.2}) and the functional integral
well-defined in the bare theory, a cutoff $s_0$ is introduced such
that all simultaneous integrals of any variables $s$ and $s^{
\prime }$ on the same chain are cut off by $| s-s^{ \prime }
| > s_0$. Let us note that the continuous chain model
(\ref{2.1}) may  be understood as a limit of discrete
self-avoiding walks, when the length of each step is decreasing
$\ell_0\to 0$ while the number of steps $N_a$ is increasing
keeping the `Gaussian surface' $S_a=N_a\ell_0^2$ fixed. The
continuous chain model (\ref{2.2}) can be mapped onto a
corresponding field theory by a Laplace transform in the Gaussian
surface variables $S_a$ to conjugate chemical potentials (``mass
variables'') $\mu_a$ \cite{Schaefer91}:
\begin{equation} \label{2.3}
\tilde{\cal Z}_f\{\mu_a\} = \int _0^{\infty} \prod_b {\bf d}S_b
\mathrm{e}^{-\mu_b S_b} {\cal Z}_f\{ S_a \} .
\end{equation}
The Laplace-transformed partition function $\tilde{\cal
Z}_f\{\mu_a\}$ can be expressed as the $m = 0$ limit of the
functional integral over vector  fields $\phi_{a}, \,
a=1,\ldots,f$ with $m$ components $\phi_a^{\alpha}, \, \alpha =
1,\ldots,m$ :
\begin{equation} \label{2.4}
\tilde{\cal Z}_f\{\mu_b\} =
  \int{\cal D}[\phi_a(r)]
   \exp[-{\cal L}\{\phi_b,\mu_b\}] \, |_{m=0}.
\end{equation}
The Landau-Ginzburg-Wilson Lagrangian $\cal L$ of $f$ interacting
fields $\phi_b$ each with $m$ components reads
\begin{equation}\label{2.5}
{\cal L}\{\phi_b,\mu_b\} = \frac{1}{2} \sum_{a=1}^{f}
\int{\rm d}^d r
\left(\mu_a\phi_a^2 + (\nabla \phi_a(r) )^2 \right)
+ \frac{1}{4!}
\sum_{a,a^{'}=1}^{f} u_{a,a'} \int {\rm d}^d r
\phi_a^2(r)\phi_{a'}^2(r) ,
\end{equation}
here $  \phi_a^2 = \sum_{\alpha = 1}^{m}( \phi_a^{\alpha} )^2 $.
The limit $m=0$ in (\ref{2.4}) can be understood as a selection
rule for the diagrams contributing to the perturbation theory
expansions which can be easily checked diagrammatically. A formal
proof of (\ref{2.4}) using the Stra\-to\-no\-vich-Hubbard
transformation to linearize terms in (\ref{2.1}) is given for the
multi-component case in \cite{Schaefer91}. The one particle
irreducible (1PI) vertex functions $\Gamma^{(L)}(q_i)$ of this
theory are defined by:
\begin{equation}\label{2.6}
 \delta\left(\sum q_i\right) \Gamma^{(L)}_{a_1...a_L}(q_i) = \int
{\rm e}^{{\rm i}q_i r_i} {\rm d} r_1 \dots  {\rm d} r_L \langle
\phi_{a_1}(r_1)\dots \phi_{a_L}(r_L) \rangle^{\cal L}_{{\rm
1PI},m=0}.
\end{equation}
The average $\langle\cdots\rangle$ in (\ref{2.6}) is understood
with respect to the Lagrangian (\ref{2.5}) keeping only those
contributions which correspond to one-particle-irreducible graphs
and which have non-vanishing tensor factors in the limit $m=0$.
The partition function $Z_{*f} \{ S_a \}$ of a polymer star
consisting of $f$ polymers of different species $1, \dots, f$
constrained to have a common end point is obtained from
(\ref{2.2}) by introducing an appropriate product of
$\delta$-functions ensuring the ``star-like'' structure. It reads:
\begin{equation}\label{2.7}
Z_{*f} \{ S_a \} = \int {\cal D} [ r_a ] \exp \left\{ -\frac{1}{k_{\rm
B}T}{\cal H}(r_a)\right\} \prod_{a=2}^f \delta^d(\vec{r}_a(0) -
\vec{r}_1(0)).
\end{equation}
The vertex part of its Laplace transformation may be defined by:
\begin{eqnarray}\label{2.8}
\lefteqn{\delta(p+\sum q_i) \Gamma^{(*f)}(p,q_1 \dots q_f) =}&&\nonumber\\
&&=
 \int {\rm e}^{{\rm i}(p r_0 + q_ir_i)} {\rm d}^d r_0 {\rm d}^d r_1 \dots
{\rm d}^d r_f  \langle \phi_1(r_0)\dots \phi_f(r_0)
\phi_1(r_1)\dots \phi_f(r_f) \rangle^{\cal L}_{{\rm 1PI},{m=0}}\, ,\quad\strut
\end{eqnarray}
where all $a_1, \dots a_f$ are distinct. The vertex function
$\Gamma^{(*f)}$ is thus defined by an insertion of the composite
operator $\prod_a\phi_a$. Its scaling properties are governed by
the scaling dimension of this operator.

In the following  we will be mainly interested in two different
cases: (i) a polymer star consisting of chains of one species with
the single interaction $u$ between them  (i.e. a {\em homogeneous
polymer star}) and (ii) a polymer star constituted by  two species
of polymers, with interactions $u_{11}$, $u_{22}$ between the
polymers of the same species and $u_{12} \, = \, u_{21}$ between
the polymers of different species (i.e. a {\em copolymer star}).
In the first case (i) one can also define $\Gamma^{(*f)}$ by the
insertion of a composite operator of traceless symmetry
\cite{Wallace75}. In the case (ii) the composite operator in
(\ref{2.8}) reduces to the product of two power-of-field operators
with appropriate symmetry $(\phi)^{f_1} (\phi^{\prime})^{f_2}$
each corresponding to a product of fields of the same `species'.
Nevertheless, the results are easily generalized to the case of
any number of polymer species.

As is well known, ultraviolet divergences occur when the vertex
functions (\ref{2.6}), (\ref{2.8}) are evaluated naively
\cite{RGbooks}. We apply renormalization group (RG) theory to make
use of the scaling symmetry of the system in the asymptotic limit
to extract the universal content and at the same time remove
divergences which occur for the evaluation of the bare functions
in this limit \cite{RGbooks}. The theory given in terms of the
initial bare variables is mapped to a renormalized theory. This is
achieved by a controlled rearrangement of the series for the
vertex functions. Several asymptotically equivalent procedures
serve to this purpose. The results reviewed in the next sections
were obtained by  two somewhat complementary approaches: zero mass
renormalization (see \cite{RGbooks} for instance) with successive
$\varepsilon$-expansion \cite{Wilson72} and the fixed dimension
massive RG approach \cite{Parisi80}. The first approach is
performed directly for the critical point. Results for critical
exponents at physically interesting dimensions $d=2$ and $d=3$ are
calculated in an $\epsilon=4-d$ expansion. The second approach
renormalizes off the critical limit but calculates the critical
exponents directly in space dimensions $d=2$, $d=3$.

Let us formulate the relations for a renormalized theory in terms
of the corresponding renormalization conditions. Though they are
different in principle for the two procedures, we may formulate
them simultaneously using the same expressions. Note that the
polymer limit of zero component fields leads to essential
simplification. Each field $\phi_a$, mass $m_a$ and coupling
$u_{aa}$ renormalizes as if the other fields were absent. First we
introduce renormalized couplings $g_{ab}$ by:
\begin{eqnarray}\label{2.9}
u_{aa} &=& \mu^{\varepsilon} Z^{-2}_{\phi_a}Z_{aa} g_{aa},
\hspace{1em} a=1,2 ,
\\ \label{2.10}
u_{12} &=& \mu^{\varepsilon} Z_{\phi_1}^{-1}Z_{\phi_2}^{-1}Z_{12}
g_{12} .
\end{eqnarray}
Here, $\mu$ is a scale parameter equal to the renormalized mass at
which the massive scheme is evaluated and sets the scale of the
external momenta in the massless scheme. The renormalization
factors $Z_{\phi_a}, Z_{ab}$ are defined as power series in the
renormalized couplings which fulfil the following RG conditions:
\begin{eqnarray}  \label{2.11}
Z_{\phi_a}(g_{aa}) \frac{\partial}{\partial k^2}
\Gamma_{aa}^{(2)}(u_{aa}(g_{aa})) &=& 1,
\\ \label{2.12}
Z^2_{\phi_a}(g_{aa}) \Gamma_{aaaa}^{(4)}(u_{aa}(g_{aa})) &=&
\mu^\varepsilon g_{aa}, \label{2.12a}\\ Z_{\phi_1}(g_{11})
Z_{\phi_2}(g_{22}) \Gamma_{1122}^{(4)}(u_{ab}(g_{ab})) &=&
\mu^\varepsilon g_{12}.
\end{eqnarray}
These formulas are applied perturbatively while the corresponding
loop integrals are evaluated for zero external momenta in the
massive approach and for external momenta at the scale of $\mu$ in
the massless approach \cite{RGbooks,FerHol97c}. In the massive
case the RG condition for the vertex function $\Gamma^{(2)}$ reads
\begin{equation}\label{2.13}
Z_{\phi_a}(g_{aa}) \Gamma_{aa}^{(2)}(u_{aa}(g_{aa}))|_{k^2=0}=
\mu^2, \hspace{1em} a=1,2.
\end{equation}
In the case of massless renormalization, the corresponding
condition reads \cite{RGbooks}:
\begin{equation}\label{2.14}
Z_{\phi_a}(g_{aa}) \Gamma_{aa}^{(2)}(u_{aa}(g_{aa}))|_{k^2=0} = 0,
\hspace{1em} a=1,2.
\end{equation}
In order to renormalize the star vertex functions we introduce
renormalization factors $Z_{* f_1,f_2}$ by
\begin{equation}\label{2.15}
Z_{\phi_1}^{f_1/2}Z_{\phi_2}^{f_2/2}Z_{* f_1,f_2} \Gamma^{(* f_1
f_2)}(u_{ab}(g_{ab})) = \mu^{\delta_{f_1+ f_2}}.
\end{equation}

With the same formalism one can also describe a star of $f$
mutually avoiding walks \cite{Duplantier88}. In this case all
interactions on the same chain $u_{aa}$ vanish and only those
$u_{ab}$ with $a \neq b$ remain. Then we define the appropriate
renormalization for the vertex function of mutually avoiding walks
(MAW):
\begin{equation}\label{2.16}
Z_{\phi_1}^{f/2}Z_{({\rm MAW} f)} \Gamma_{{\rm
MAW}}^{*f}(u_{12}(g_{ab})) = \mu^{\delta_f}.
\end{equation}
The powers of $\mu$ absorb the engineering dimensions of the bare
vertex functions. These are given by $\delta_f = f(\varepsilon/2
-1)+ 4 - \ \varepsilon$. The renormalized couplings $g_{ab}$
defined by the relations (\ref{2.9}), (\ref{2.10}) depend on the
scale parameter $\mu$. By their dependence on $g_{ab}$ also the
renormalization $Z$-factors implicitly depend on $\mu$. This
dependence is expressed by the RG functions defined by the
following relations:
\begin{eqnarray} \label{2.18}
\mu \frac{\rm d}{{\rm d}\mu} g_{ab} &=& \beta_{ab}(g_{a'b'}),
\\ \label{2.19}
\mu \frac{\rm d}{{\rm d}\mu} \ln Z_{\phi_a} &=&
\eta_{\phi_a}(g_{aa}),
\\  \label{2.20}
 \mu \frac{\rm
d}{{\rm d}\mu} \ln Z_{*f_1f_2} &=& \eta_{*f_1f_2}(g_{ab}),
\label{2.21}\\ \mu \frac{\rm d}{{\rm d}\mu} \ln Z_{{\rm MAW}f} &=&
\eta^{{\rm MAW}}_f(g_{ab}).
\end{eqnarray}
The function $\eta_{\phi_a}$ describes the pair correlation
critical exponent, while the functions $\eta_{*f_1f_2}$ and
$\eta^{{\rm MAW}}_f(g_{ab})$ define the set of exponents for
copolymer stars and the stars of mutually avoiding walks. Note that
$Z_{*20}$ renormalizes the vertex function with a $\phi^2$
insertion which coincides with $\Gamma^{(*20)}$. Consequently, the
usually defined correlation length critical exponent $\nu$ is
expressed in terms of functions  $\eta_{*20}$ and $\eta_{\phi}$.

Expressions for the $\beta$ and $\eta$ functions will be discussed
in the next sections together with a study of the RG flow and the
fixed points of the theory.


\section{Homogeneous star polymer in three dimensions} \label{III}

First we give numerical results for a {\em homogeneous polymer
star} constituted by $f$ polymer chains of the same species. The
scaling of a star is governed by the laws~(\ref{1.4}). For one
species of polymers ($a,b=1$), corresponding RG function
(\ref{2.18})--(\ref{2.20}) depend on a single coupling $g_{ab}=g$
and reduce to a single functions
$\beta_{ab}(g_{a'b'})\equiv\beta(g)$,
$\eta_{\phi_a}(g_{aa})\equiv\eta_{\phi}(g)$ and a set of functions
$\eta_{*f_1f_2}(g_{ab})\equiv \eta_{*(f_1+f_2)}(g)$ with
$f_1+f_2=f$. Here, the function $\beta(g)$ is a standard
$\beta$-function of the $O(m=0)$-symmetric theory. In the
infrared-stable fixed point $g^*$:
\begin{equation}\label{3.1}
\beta(g^*)=0, \hspace{3em} \frac{{\rm d}\beta(g)}{{\rm d} g}
|_{g=g^*}>0;
\end{equation}
the function $\eta_{\phi}(g)$ gives the pair correlation function
critical exponent $\eta$:
\begin{equation}\label{3.2}
\eta=\eta(g^*),
\end{equation}
whereas  the functions $\eta_{*f)}(g)$ give the scaling exponents
$\eta_f$ (\ref{1.4}):
\begin{equation}\label{3.3}
\eta_f=\eta_{*f}(g^*).
\end{equation}
The RG functions (\ref{3.1})--(\ref{3.3}) are obtained in the form
of perturbation theory series. In the massive RG scheme, these are the
series in a renormalized coupling $g$ for fixed space dimension
$d$ whereas the zero-mass renormalization expansion is performed
both in $g$ and in $\varepsilon=4-d$. Increasing the order of
perturbation theory corresponds to increasing the number of loops
when the expansion is written in terms of Feynman diagrams. That
is why sometimes the expansions are classified in successive
approximations in number of loops. The functions (\ref{3.1}),
(\ref{3.2}) are the standard RG functions of the $O(m=0)$
symmetric theory. They are known by now in high orders of
perturbation theory, whereas the functions (\ref{3.3}) have been
obtained only up to three loops. In the field theory they are related
to the anomalous dimension of composite operators of traceless
symmetry \cite{Wallace75}. In polymer language the star exponents
were originally calculated to order $\varepsilon$ in
\cite{Miyake83}, the $\varepsilon^2$ results have been obtained in
\cite{Duplantier89,Ohno89}, whereas $\varepsilon^3$ expansions are
given in \cite{Schaefer92}. The $\varepsilon$-expansion for the
exponent $\eta_f$ reads \cite{Wallace75,Schaefer92}:
\begin{equation}\label{3.4}
\eta_f=-\frac{\varepsilon}{8}f(f-1) \Big[1-
\frac{\varepsilon}{32}(8f-25)+
 \frac{\varepsilon^2}{64}[(28f-89)\zeta(3)+8f^2-49f+\frac{577}{8}]+
 O(\varepsilon^3) \Big].
\end{equation}

In the massive RG scheme, the three-loop expansions for the $d=3$
star exponents were obtained in the form of pseudo-$\varepsilon$
expansion in \cite{FerHol9596}. However, as already noted
above, the perturbation theory expansions of the renormalized
field theory have zero radius of convergence and are asymptotic at
best. Therefore, in order to make numerical estimates for the
expansions one should apply an appropriate resummation procedure.
Here, we will not describe the different ways of resummation which
were applied to the series for the star exponents referring rather to
the papers \cite{Schaefer92,FerHol97c,FerHol9596,FerHol99} where
they are given in detail. For the purpose of the present review
we just mention that a Borel transform was applied to get rid of
the (conjectured) factorial divergence of the perturbation theory
terms \cite{Hardy}. Subsequently, the expansion for the Borel
transformed function was refined by a conformal mapping from the
initial variable defined on the cut-plane to a disk keeping the origin
invariant. Another scheme of resummation
exploits the  analytic continuation of the series for the Borel
transformed function by means of Pad\'e approximant (Pad\'e-Borel
resummation).

\begin{table}
 \caption {\label{tab1}  Values of the star exponent $\gamma_{f}$
obtained in three-dimensional theory  (columns 2,3,4,5) in
comparison with the results of $\varepsilon$-expansion (columns
6,7,8) and Monte-Carlo simulations (column 9). See the text for a
full description.}
\vspace{2ex}

\begin{center}
 \begin{tabular}{|r|rrrr|rrr|l|}
 \hline
 $f$ & \multicolumn{4}{c|} {pseudo-$\varepsilon$ expansion
 \cite{FerHol9596}}  &
\multicolumn {3}{c|} {$\varepsilon$-expansion } & MC   \\ &
\multicolumn{2}{c} {Pad\'e-Borel} & \multicolumn{2}{c|}
{conf.mapping} & \multicolumn {3}{c|} {\cite{Schaefer92} } &
 \cite{Batoulis89,Barret87}  \\
 \hline
 3 &  1.06 &  1.05 & 1.06 &1.06 &  1.05 &  1.05 &  1.07 &  1.09 \\
 4 &  0.86 &  0.86 & .86  &.83  &  0.84 &  0.83 &  0.85 &  0.88 \\
 5 &  0.61 &  0.61 & .58  &.56  &  0.53 &  0.52 &  0.55 &  0.57 \\
 6 &  0.32 &  0.32 & .24  &.22  &  0.14 &  0.18 &       &  0.16 \\
 7 & --0.02 & --0.01 & --.17 &--.17 & --0.33 & --0.20 &       &        \\
 8 & --0.40 & --0.36 & --.63 &--.62 & --0.88 & --0.60 &       & (--0.99,
 --0.30) \\
 9 & --0.80 & --0.72 & --1.14&--1.11& --1.51 & --1.01 &       &       \\
 \hline
 \end{tabular}
 \\
 \end{center}
 \end{table}

In table \ref{tab1} we give the numerical results for the star
exponents $\gamma_f$ as obtained by the fixed $d=3$ RG technique
and in the $\varepsilon$-expansion in comparison with the data of
the numerical simulation. First, we give the values of the
exponent $\gamma_f$ obtained in the fixed $d=3$ technique by
Pad\'e-Borel resummation \cite{FerHol9596}: the second column
contains the value of the $\gamma_f$ obtained directly from the
resummation of the series for $\gamma_f$, while the third column
gives the $\gamma_f$ based on the resummed series for the star
exponents $x^{'}_{f}$. These are connected with the $\gamma_f$ via
the scaling relation:
\begin{equation}\label{3.5}
\gamma_{f}= 1-\nu x^{'}_{f}+[\nu(2-\eta/2)-1]f
\end{equation}
substituting the well-known values of the exponents
$\nu(d=3)=0.588$, $\eta(d=3)=0.027$. Columns 4 and 5 give
$\gamma_f$ obtained by the resummation using the conformal mapping
technique \cite{FerHol9596}: the resummation of the series
$\gamma_f(\tau)$ (fourth column) and the resummation of the series
$x^{'}_{f}(\tau)$ (fifth column). The next columns show the
results obtained by the $\varepsilon^{3}$-expansion using: simple
Pad\`{e} approximation (the 6th column) and Pad\`{e}-Borel
analysis neglecting or exploiting the exact results for $d=2$ (the 7th
and 8th columns, respectively) \cite{Schaefer92}. The last column
contains Monte-Carlo data \cite{Batoulis89,Barret87}. For low
numbers of arms $F\leqslant 5$ the results of the different approaches
agree reasonably well and are also close to the values obtained by
MC simulation.

The data collected in table \ref{tab1} was obtained using two
different renormalization schemes as well as different procedures
for the resummation of the resulting asymptotic series. Thus, table
\ref{tab1} gives a test for the stability of the results
under the changes of the calculational scheme.  Obviously, for higher
numbers $f>5$ of arms, the coincidence of the results is no longer
good. The main reason for this is that calculating the exponents,
combinatorial factors lead to an expansion in $f\varepsilon$ for
the $\varepsilon$ expansion and of $fg$ when directly expanding in
a renormalized coupling $g$. For such large values of the
expansion parameter, even resummation of the series fails. For
larger numbers of arms, other approaches to the theory of polymer
stars, like a self consistent field approximation, might be more
useful.


\section{Copolymers and copolymer stars} \label{IV}

\begin{wrapfigure}{i}{0.5\textwidth}
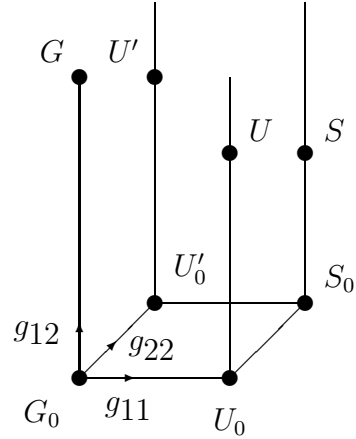

\begin{center}
\input fig1.pic
\end{center}
\caption{Fixed points (FPs) of ternary polymer solution. The
trivial FPs $G_0$, $U_0$, $U_0^{\prime}$, $S_0$ correspond to a
vanishing mutual interaction. The non-trivial FPs $G$, $U$,
$U^{\prime}$, $S$ correspond to a non-vanishing mutual interaction
($g_{12} \neq 0$).  } \label{fig1}
\end{wrapfigure}

Now let us pass to the case of a polymer star constituted by two
species of polymers. It is described by the Lagrangian (\ref{2.5})
with interactions $u_{11}$, $u_{22}$ between the polymers of the
same species and $u_{12} \, = \, u_{21}$ between the polymers of
different species. Such a Lagrangian is used to describe the system of
polymers of two species immersed in a solvent: the so-called
ternary solution \cite{ternary}. A comprehensive analysis of the
fixed point behaviour of a ternary solution  was given in
\cite{Schaefer91} where the RG flow of the theory was calculated
within the massless renormalization and is known by now to the
third loop order of the $\varepsilon$-expansion. Note, that for
the diagonal coupling $g_{aa}$ the corresponding expressions are
also found in the polymer limit $m=0$ of the $O(m)$-symmetric
$\phi^4$ model. They are known in even higher orders of
perturbation theory. In the massive RG approach, the corresponding
expressions were obtained in \cite{FerHol97c}. The equations
for the fixed points (FP) P$(\{g^*_{11}, g^*_{22}, g^*_{12}\})$ of
the $\beta$-functions read:
\begin{eqnarray}
\beta_{g_{aa}} (g^*_{aa})&=& 0, \hspace{2em} a=1,2, \nonumber
\\ \beta_{g_{12}}(g^*_{11}, g^*_{22}, g^*_{12}) &=&
0. \label{4.1}
\end{eqnarray}
As is well known, the first equation has two solutions
$g^*_{aa}=0,g^*_{\rm S}$. For the second equation, one finds a total of 8
FPs depending on the choice for $g^*_{aa}$. The trivial FPs are
${\rm G}_0(0,0,0), {\rm U}_0(g^*_{\rm S},0,0), {\rm
U'}_0(0,g^*_{\rm S},0), {\rm S}_0(g^*_{\rm S},g^*_{\rm S},0)$, all
corresponding to $g^*_{12}=0$. The non-trivial FPs are found as
${\rm G}(0,0,g^*_{\rm G})$, ${\rm U}(g^*_{\rm S},0,g^*_{\rm U})$,
${\rm U'}(0,g^*_{\rm S},g^*_{\rm U})$, and ${\rm S}(g^*_{\rm
S},g^*_{\rm S},g^*_{\rm S})$. In the three dimensional space of
couplings $g_{11}, g_{22}, g_{12}$ these FPs  are placed at the
corners of a cube that is deformed in the $g_{12}$ direction  (see
figure~\ref{fig1}).

Looking for the stability of the above described fixed points one
finds that only the fixed point {\it S} is stable
\cite{Schaefer91}. In the excluded volume limit of infinitely long
chains, the behaviour of a system of two polymer species is thus
described by the same scaling laws as a solution of only one
polymer species. Nevertheless, taking into account that real
polymer chains are not infinitely long, one may also find crossover
phenomena which are governed by the unstable fixed points. Knowing
the complete RG flow, allows one to describe crossover phenomena in
the whole accessible region \cite{Schaefer91}. However, for the
purpose of our review we are interested only in the values of the
fixed points and in the properties of the star vertex functions at
these points.

Let us define the asymptotic values of the copolymer star
exponents and the mutually avoiding walk (MAW) star exponents by:
\begin{eqnarray} \label{4.2}
\eta^{\mathrm{S}}_{f_1f_2} &=& \eta_{*f_1f_2}(g_{ab})|_S ,
\\ \label{4.3}
\eta^{\mathrm{G}}_{f_1f_2} &=& \eta_{*f_1f_2}(g_{ab})|_G ,
\\ \label{4.4}
\eta^{\mathrm{U}}_{f_1f_2} &=& \eta_{*f_1f_2}(g_{ab})|_U=
\eta_{*f_2f_1}(g_{ab})|_{U^{\prime}},
\\ \label{4.5}
\eta^{{\rm MAW}}_f &=& \eta^{{\rm MAW}}_f(g_{ab})|_G.
\end{eqnarray}
Taking into account the nature of the fixed points, where the
exponents (\ref{4.1})--(\ref{4.5}) are defined, one arrives at the
following physical interpretation:
\begin{itemize}
\item the exponent $\eta^{\mathrm{G}}_{f_1f_2}$ describes a star of $f_1$ random
walks of the 1st type and $f_2$ random walks of  the 2nd type all
of which are not selfavoiding; only the chains of the 1st type
avoid those of the  2nd type and vice versa;
\item the exponent $\eta^{\mathrm{U}}_{f_1f_2}$ describes a star of $f_1$ self and
mutually avoiding polymer chains walks of the 1st type and $f_2$
random walks which only avoid those of the 1st type;
\item the exponent $\eta^{\mathrm{MAW}}_{f_1f_2}$ describes a star of $f_1+f_2$ random
walks all of which are not selfavoiding; however, each chain avoids
any other chain of the star;
\item exponent $\eta^{\mathrm{S}}_{f_1f_2}$ describes a star of $f_1+f_2$ self
  and mutually avoiding chains. The exponent can also be expressed by
  $\eta^{\mathrm{S}}_{f_1f_2}=\eta^{\mathrm{U}}_{f_1+f_2,0}$ (This relation is broken for
  the exact results known in $d=2$ where a finite effect remains in
  the zero walk limit). This situation has already been discussed in
  the previous section.
\end{itemize}

Starting from the expressions for the fixed points (available both
in the form of a three-loop $\varepsilon$ expansion
\cite{Schaefer91} or in the form of pseudo-$\varepsilon$ expansion
at fixed $d$ \cite{FerHol97c}) and the relations (\ref{4.2})--(\ref{4.5}),
one can find the series for the star exponents. In
the $\varepsilon$-expansion the following expansions for
$\eta_{f_1f_2}$ are obtained \cite{FerHol97c}:
\begin{eqnarray}
\eta^{\mathrm{G}}_{f_1f_2} (\varepsilon) &=& - f_1 \, f_2 \frac
{\varepsilon}{2}+ f_1 \, f_2 \,\Big ( f_2 - 3 + f_1 \Big )
\frac{\varepsilon^2}{8}
\nonumber \\
&& {}- f_1 \, f_2 \Big ( f_2 - 3 + f_1 \Big )
\Big ( f_1 +f_2 + 3 \, \zeta (3)- 3 \Big )\frac
{\varepsilon^3}{16}, \label{4.6}
\\ [2ex]
\eta^{\mathrm{U}}_{f_1f_2}(\varepsilon) &=&
{ f_1}\Big (1-{ f_1}- 3{ f_2}\Big )\frac {\varepsilon}{8}
\nonumber \\
&&{}+ { f_1}\Big (25-33{ f_1}+
8{{ f_1}}^{2}- 91{ f_2}+ 42{ f_1}{ f_2}+
18{{ f_2}}^{2}\Big) \frac {\varepsilon^2}{256}
\nonumber \\
&& {}+ {
f_1}\Big (577- 969{ f_1}+ 456{{ f_1}}^{2}- 64{{
f_1}}^{3}-2463{ f_2}+ 2290{ f_1}{ f_2}
\nonumber \\
&&\quad{}-
492{{ f_1}}^{2}{ f_2}+ 1050{{ f_2}}^{2}- 504{
f_1}{{ f_2}}^{2}- 108{{ f_2}}^{3}- 712\zeta (3)+ 936{
f_1}\zeta (3)
\nonumber \\
&&\quad{}-
224{{ f_1}}^{2}\zeta (3)+
2652{ f_2}\zeta (3)- 1188{ f_1}{ f_2}\zeta (3)- 540{{
f_2}}^{2}\zeta (3)\Big ) \frac {\varepsilon^3}{4096},
\nonumber\\ [-0.5ex]
 \label{4.7}
\\ [2ex]
\eta^{\mathrm{MAW}}_{f_1f_2}(\varepsilon) &=& - ({ f_1}- 1 ){ f_1}
\frac {\varepsilon}{4}+ { f_1} ({ f_1}- 1 ) (2{ f_1}- 5 )
\frac {\varepsilon^2}{16}
\nonumber \\
&& {}-({ f_1}-1 ){ f_1} (4{{ f_1}}^{2}-20{ f_1}+ 8{
f_1}\zeta (3)- 19\zeta (3)+ 25 ) \frac {\varepsilon^3}{32}.
\label{4.8}
\end{eqnarray}
Here $\zeta(3)\simeq 1.202$ is the Riemann $\zeta$-function. The
above formulas reproduce the 3rd order calculations
\cite{Schaefer92} of the scaling exponents of homogeneous polymer
stars $\eta_f=\eta^{\rm
  U}_{f,0}$ given in formula (\ref{4.4}) of the previous section. The
corresponding pseudo-$\varepsilon$ expansion for the exponents
$\eta_{f_1f_2}$ obtained in the massive scheme may be found in
\cite{FerHol97c}. We do not give it here explicitly although in
what follows below we will give numerical values of the exponents
obtained in both approaches. It has been pointed out in
\cite{Cates} that for the exponent $\eta_{12}^{\mathrm{G}}$ an exact estimate
equal to our first order contribution may be found. It is indeed
remarkable that all higher order contributions to $\eta_{12}^{\mathrm{G}}$
vanish in both approaches, $\eta_{12}^{\mathrm{G}}=-\varepsilon$ being an
exact result \cite{Semenov}.

\begin{table}[h]
\caption{ \label{tab2} Values of the copolymer star exponent
$\eta_{f_1f_2}^{\mathrm{G}}$ for $d=3$ obtained by the
$\varepsilon$-expansion ($\varepsilon$) and by fixed dimension
technique ($3d$). }
\vspace{2ex}

\tabcolsep=1mm
\begin{center}
\begin{tabular}{|c|rr|rr|rr|rr|rr|rr|}
\hline
$f_1$ & \multicolumn{2}{c|}{$1$}& \multicolumn{2}{c|}{$2$}&
\multicolumn{2}{c|}{$3$}& \multicolumn{2}{c|}{$4$}&
\multicolumn{2}{c|}{$5$}& \multicolumn{2}{c|}{$6$}\\ $f_2$ &
$\varepsilon$ & $3d$ & $\varepsilon$ & $3d$ & $\varepsilon$ & $3d$
& $\varepsilon$ & $3d$ & $\varepsilon$ & $3d$ & $\varepsilon$ &
$3d$ \\ \hline 1 &
  --0.56 &  --0.58 &
  --1.00 &  --1.00 &
  --1.33 &  --1.35 &
  --1.63 &  --1.69 &
  --1.88 &  --1.98 &
  --2.10 &  --2.24 \\
2 & & &
  --1.77 &  --1.81 &
  --2.45 &  --2.53 &
  --3.01 &  --3.17 &
  --3.51 &  --3.75 &
  --3.95 &  --4.28 \\
3 & & & & &
 --3.38 &  --3.57 &
 --4.21 &  --4.50 &
 --4.94 &  --5.36 &
 --5.62 &  --6.15 \\
4 & & & & & & &
 --5.27 &  --5.71 &
 --6.24 &  --6.84 &
 --7.12 &  --7.90 \\
5 & & & & & & & & &
 --7.42 &  --8.24 &
 --8.50 &  --9.54 \\
6 & & & & & & & & & & &
 --9.78 &  --11.07\\
\hline
\end{tabular}
\end{center}
\end{table}

\begin{table}[h]
\caption{ \label{tab3}The  values of the copolymer star exponent
$\eta_{f_1f_2}^{\mathrm{U}}$ at $d=3$ obtained by the
$\varepsilon$-expansion ($\varepsilon$) and by fixed dimension
technique ($3d$). }
\vspace{2ex}

\tabcolsep=1mm
\begin{center}
\begin{tabular}{|c|rr|rr|rr|rr|rr|rr|}
\hline
$f_1$ & \multicolumn{2}{c|}{$1$}& \multicolumn{2}{c|}{$2$}&
\multicolumn{2}{c|}{$3$}& \multicolumn{2}{c|}{$4$}&
\multicolumn{2}{c|}{$5$}& \multicolumn{2}{c|}{$6$}\\ $f_2$ &
$\varepsilon$ & $3d$ & $\varepsilon$ & $3d$ & $\varepsilon$ & $3d$
& $\varepsilon$ & $3d$ & $\varepsilon$ & $3d$ & $\varepsilon$ &
$3d$ \\ \hline 0 &
    0 &  0    &
--0.28 & --0.28 & --0.75 & --0.76 & --1.36 & --1.38 & --2.07 & --2.14 &
--2.88 & --3.01 \\ 1 & --0.43 &  --0.45 & --0.98 &  --0.98 & --1.64 &
--1.67 & --2.39 &  --2.47 & --3.21 &  --3.38 & --4.11 &  --4.40 \\ 2 &
--0.79 &  --0.81 & --1.58 &  --1.60 & --2.44 &  --2.52 & --3.33 &  --3.50
& --4.28 &  --4.57 & --5.29 &  --5.73 \\ 3 & --1.09 &  --1.09 & --2.13 &
--2.19 & --3.16 &  --3.30 & --4.20 &  --4.48 & --5.28 &  --5.71 & --6.41 &
--7.03 \\ 4 & --1.35 &  --1.37 & --2.61 &  --2.71 & --3.82 &  --4.04 &
--5.02 &  --5.40 & --6.24 &  --6.81 & --7.48 &  --8.28 \\ 5 & --1.60 &
--1.64 & --3.05 &  --3.21 & --4.44 &  --4.75 & --5.80 &  --6.30 & --7.15 &
--7.89 & --8.51 &  --9.50 \\ 6 & --1.81 &  --1.89 & --3.46 &  --3.68 &
--5.01 &  --5.42 & --6.53 &  --7.15 & --8.02 &  --8.92 & --9.50 & --10.69\\
\hline
\end{tabular}
\end{center}
\end{table}

With these exponents one can describe the scaling behaviour of
polymer stars and networks of two components, generalizing the
relation for single component networks \cite{Duplantier89}. In the
notation of (\ref{1.4}) one finds for the number of configurations
of a network $\cal G$ of $F_1$ and $F_2$ chains of species $1$ and
$2$
\begin{equation}
\label{4.9} {\cal Z}_{\cal G} \sim (R/\ell)^{\eta_{\cal G}
-F_1\eta_{20}-F_2\eta_{02}} \mbox{, \, \,with } \eta_{\cal G} =
-\mathrm{d}L + \sum_{f_1+f_2\geqslant 1} N_{f_1f_2}\eta_{f_1f_2},
\end{equation}
where $L$ is the number of loops and $N_{f_1f_2}$ is the number of
vertices with $f_1$ and $f_2$ arms of species $1$ and $2$ in the
network $\cal G$. To receive an appropriate scaling law we assume
the network to be built of chains which for both species will have
the same coil radius $R$ when isolated.

\begin{table}[h]
\caption{ \label{tab4} The values of the $\eta^{\mathrm{MAW}}_f$ exponents
of stars of mutually avoiding walks for $d=3$, $d=2$ obtained by
the $\varepsilon$-expansion ($\varepsilon$) and by the fixed
dimension technique ($3d$, $2d$). The last column gives the exact
conjecture for $d=2$ \protect\cite{Duplantier88}.}
\vspace{2ex}

\begin{center}
\begin{tabular}{|l|ll|lll|}
\hline
&\multicolumn{2}{c|} {$d=3$}&\multicolumn{3}{c|} {$d=2$} \\ $f$ &
$\varepsilon$ & $3d$ & $\varepsilon$ & $2d$ & exact \\ \hline 1 &
0     &   0    &   0     &   0     & --.250      \\ 2  & --.56 &
--.56  & --1.20   & --1.19   & --1.250     \\ 3  & --1.38  &  --1.36 &
--2.71   & --2.60   & --2.916(6)  \\ 4  & --2.36  &  --2.34 & --4.36 &
--4.07   & --5.250     \\ 5  & --3.43  &  --3.43 & --6.04   & --5.61 &
--8.250     \\ 6  & --4.58  &  --4.64 & --7.78   & --7.17   &
--11.916(6)\\
\hline
\end{tabular}
\end{center}
\end{table}

In order to calculate numerical values for the exponents
$\eta_{f_1f_2}^{\rm G}$, $\eta_{f_1f_2}^{\rm U}$ and
$\eta^{\mathrm{MAW}}_f$ in
\cite{FerHol97c,FerHol97a,FerHol97b,FerHol98} the Borel
resummation refined by conformal mapping was applied to the series
(\ref{4.6})--(\ref{4.8}) as well as to the appropriate series in
pseudo-$\varepsilon$ expansion. The resummation showed that the
two schemes yield consistent numerical estimates. Our tables
\ref{tab2}, \ref{tab3}, and \ref{tab4} list the results.

Comparing the numerical values listed in the above tables it is
convincing that the two approaches and the different resummation
procedures all lead to the results which lie within a bandwidth of
consistency, which is broadening for larger values of number of
chains $f_1,f_2 >1$. This is not surprising. One can see, e.g. from the
formulas  (\ref{4.6})--(\ref{4.8}), that the expansion parameters
are multiplied by $f_1$ and $f_2$. Rather, it is remarkable that
even for a total number of chains of the order of 10 (see tables
\ref{tab2}, \ref{tab3}) one still receives the results which are
comparable to each other.

In the next three sections we will briefly review several
phenomena where star exponents become important.

\section{Colloids with polymer stars: The interaction}\label{VII}

Let us now consider the effective interaction between the cores of
two star polymers at a small distance (small on the scale of the
size $R_g$ of the star).  Let us analyse the general case of two
stars of different functionalities $f_1$ and $f_2$. The power law
for the partition sum ${\cal Z}^{(2)}_{f_1f_2}$ of such two star
polymers at a distance $r$ \cite{Duplantier89},
\begin{equation}\label{7.1}
{\cal Z}^{(2)}_{f_1f_2}(r)\sim r^{\Theta^{(2)}_{f_1f_2}},
\end{equation}
is governed by the contact exponent $\Theta^{(2)}_{f_1f_2}$. As we
have seen in the section \ref{II} in the formalism of the $m=0$
component model, the core of a star polymer corresponds to a local
product of $f$ spin fields $\phi_1({\bf x})\cdots\phi_f({\bf x})$,
each representing the endpoint of one polymer chain. The
probability of approach of the two cores of the star polymers at
a small distance $r$ in these terms is described by a short distance
expansion for the composite fields. The short distance expansion
provides the scaling relations between the exponents
$\Theta_{f_1f_2}$ and the star exponents $\gamma_f$ (or $\eta_f$):
\begin{eqnarray}\label{7.2}
\nu\Theta_{f_1f_2} &=&
(\gamma_{f_1}-1)+(\gamma_{f_2}-1)-(\gamma_{f_1+f_2}-1)\,,
\nonumber\\ \Theta_{f_1f_2} &=&
\eta_{f_1}+\eta_{f_2}-\eta_{f_1+f_2}\,.
\end{eqnarray}
The mean force $F^{(2)}_{f_1f_2}(r)$ between two star polymers at
a short distance $r$ is easily derived from the effective potential
$V^{\rm eff}_{f_1f_2}(r)=\log {\cal Z}^{(2)}_{f_1f_2}(r)$ as
\begin{equation}\label{7.3}
 F^{(2)}_{f_1f_2}(r)=
\frac{\Theta^{(2)}_{f_1f_2}}{r}\,, \mbox{ with
}\,\Theta^{(2)}_{ff}  \approx\frac{5}{18}f^{3/2} .
\end{equation}
The factor $5/18$ is found by matching the cone approximation for
$\Theta^{(2)}_{ff}$ (see formula (\ref{1.5})) to the known values
of the contact exponents for $f=1,2$ \cite{Likos}. This matching,
in turn, proposes an approximate value for the $\eta_f$ exponents,
\begin{equation}\label{7.4}
 \eta_f\approx -\frac{5}{18} (2^{3/2}-2)f^{3/2} -f\eta_1,
\end{equation}
where $-f\eta_1$ is introduced for consistency with
 the exact result $\eta_1=0$.
This assumption nicely reproduces the contact exponents as derived
from 3-loop perturbation theory \cite{FerHol97c} as is displayed
in table \ref{tab5} taken from the \cite{FerHol01a}.

\begin{table}
\caption { \label{tab5} The prefactor $\Theta_{f_1f_2}$ of the
force between two star polymers at a short distance calculated in
non-resummed 1-loop and resummed 3-loop RG analysis in comparison
to the result of the cone approximation.}
\vspace{2ex}

\begin{center}
\begin{tabular}{|l|r|r|r|r|r|l|}
\hline $\Theta_{f_1f_2}$ & $f_2=1$ & $f_2=2$ & $f_2=3$ & $f_2=4$ &
$f_2=5$ & app.\\ \hline $f_1=1$           & --.21    &   --.42  &
--.63  &  --.85   &  --1.06  & 1-loop\\ $f_1=1$           & --.28    &
--.48  &   --.62  &  --.76   &   --.87  & 3-loop\\ $f_1=1$           &
--.27    &   --.45  &   --.60  &  --.73   &   --.84 & cone\\
&&&&&&\\
$f_1=2$           &         &   --.85  &  --1.27  & --1.70   & &
1-loop\\ $f_1=2$           &         &   --.82  &  --1.10  & --1.35 &
& 3-loop\\ $f_1=2$           &         &   --.78  & --1.05  & --1.29
&        &cone\\
&&&&&&\\
$f_1=3$           &         &         &  --1.91  &         & &
1-loop\\ $f_1=3$           &         &         &  --1.49  & & &
3-loop\\ $f_1=3$           &         &         & --1.44  & &
&cone\\ \hline
\end{tabular}
\end{center}
\end{table}


In table \ref{tab5} we use the approximate values of $\eta_f$  to
calculate the cone estimation of the contact exponents and compare
these with the corresponding values of a renormalization group
calculation. A perturbation series is taken in terms of an
$\varepsilon$-expansion. The result labelled `1-loop' corresponds
to optimal truncation of the series by simply inserting
$\varepsilon=1$ or $d=3$ in the first order term of the expansion.
The 3-loop result includes a resummation procedure that takes into
account the asymptotic nature of the series \cite{FerHol97c}. The
large $f$ result corresponds to the cone approximation~
(\ref{1.5}).

The above data convincingly shows that the large $f$
approximation for the short distance force between two star
polymers can be consistently fitted to the results of perturbation
theory for low values of the functionality $f$ of the stars. Note,
that increasing the number of chains $f$, the polymer stars
interpolate between the properties of linear polymers and
polymeric micelles \cite{Gast}.
 This approach is general enough to
describe the interaction between two stars of different
functionalities $f_1$ and $f_2$. This is essential in extending
the theory of colloidal solutions of star polymers to general
polydispersity in $f$ as it appears naturally in any real
experiment.

\section{Brownian motion near an absorbing polymer star}
\label{V}

\begin{wrapfigure}{i}{0.5\textwidth}
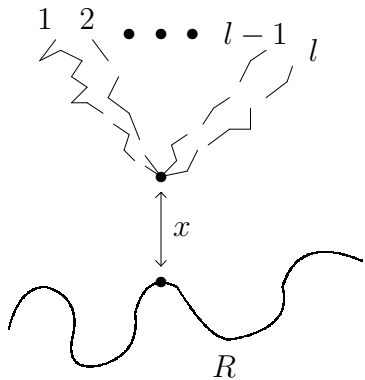

\begin{center}
\input fig2.pic
\end{center}
\caption{ \label{fig2} Star of $l$ random walks at distance $x$
from an absorbing polymer of size $R$.}
\end{wrapfigure}

Now, let us turn to another physical phenomenon that is described
in terms of the star exponents. To this end we consider a model
proposed by Cates and Witten \cite{Cates} to describe the flow of
diffusing particles near an absorbing object, which may be a
polymer chain ( = a selfavoiding walk (SAW)) or a random walk
(RW). The flux of particles on the absorbing chain is defined by
the density of incoming particles close to the absorber
(see figure~\ref{fig2}). Taken $\rho(r)$ to be the probability density of
incoming particles, the problem is to solve the steady-state
diffusion equation (Laplace equation)
\begin{equation}\label{5.1}
\nabla^2 \rho(r) = 0
\end{equation}
with boundary conditions for a given absorbing polymer
\begin{eqnarray}\nonumber
\rho(r)&=&0 \hspace{5em} \mbox{on the surface of the absorber},
\\ \label{5.2}
\rho(r)&=&\rho_{\infty}={\rm const} \hspace{5em} \mbox{ for } \quad |r|
\to \infty .
\end{eqnarray}
One of the motivations for introducing this model was to gain insight for
the problem of diffusion limited aggregation (DLA)
\cite{Witten81}. The latter phenomenon is much more complicated
because of the fact that for DLA the boundary conditions are given
on the surface of the growing aggregate itself. The process
described by equations (\ref{5.1}), (\ref{5.2}) may be rather
considered as a diffusion limited catalysis, when particles of one
type interact with a {\it prescribed} fractal (a catalyzing
polymer) and transform into the other type \cite{Oshanin} (see
section \ref{VI} for more details).

Let the absorbing polymer of size $R$ be chosen from the well
defined ensemble of SAWs. Then one may introduce exponents that
govern the scaling of the moments of the field $\rho(r)$ close to
the surface of the absorber (see figure~\ref{fig2}). For distances
$R\gg r>a$ ($a$ being a cut off) the averaged moments $\langle\rho(r)^n\rangle$
are expected to scale as \cite{Cates}
\begin{equation} \label{5.3}
\langle\rho(r)^n\rangle \sim (R/r)^{-\lambda(n)},
\end{equation}
where $\langle\dots\rangle$ denotes the average over the ensemble of polymers.
Taking that the flux $\phi(x)$ onto any randomly chosen point $x$
of the absorber is proportional to the field $\rho(r)$ at a point
that is as close as a cut off length $a$ from the absorber, one
finds for the averaged moments of the flux:
\begin{equation} \label{5.4}
\langle\phi^n\rangle \sim (R/a)^{-\lambda(n)}.
\end{equation}

There is a natural way to associate the exponents $\lambda(n)$
(\ref{5.3}), (\ref{5.4}) with the copolymer star exponents
$\eta_{f_1f_2}$ (\ref{4.3}), (\ref{4.4}). The central idea that
allows one to calculate the properties of the solutions of
equation (\ref{5.1}) with the boundary conditions (\ref{5.2}) is
that a path integral representation both for the field $\rho(r)$
and for the absorber (being a RW or a SAW) is possible
\cite{Cates,Oshanin}.  In terms of the path integral solution of
the Laplace equation, one finds that $\rho(r)$ at point $r$ near
the absorber is proportional to the number of RW that end at point
$r$ and avoid the absorber. The $n$th power of this field is
proportional to the $n$th power of the above mentioned number,
i.e. it is defined by the partition function of a star polymer
with $n$ arms (\ref{2.7}) (the latter is shown by dashed lines in
figure~\ref{fig2}). Furthermore, introducing the mutual avoidance
conditions between the ``$n$-arm star'' and the 2-arm polymer
(representing the absorber), one has to calculate the partition
function of a co-polymer star consisting of chains of two
different species that avoid each other.  These correspond to the
trajectories of diffusing particles (being RW) and the absorbing
polymer (which is chosen to be a RW or SAW).  Making use of the
theory of copolymer stars and networks \cite{FerHol97c}, one may
relate \cite{FerHol99} the spectrum of exponents (\ref{5.3}),
(\ref{5.4}) to the exponents that define the scaling properties of
co-polymer stars. In particular, as we have seen in the section
\ref{IV} for a co-polymer star consisting of $f_1$ chains of
species $1$ and $f_2$ chains of species $2$, the scaling of the
number of configurations $Z_*$ is governed by the exponents
$\eta_{f_1f_2}$ (\ref{4.9}). By means of a short-chain expansion
\cite{Ferber97} it is clear that only the (smaller) length scale
of the absorbing polymer remains and the set of exponents
$\eta_{f_1f_2}$ can be related to the exponents $\lambda(n)$
(\ref{5.3}) that govern the scaling of the $n$th moment of the
flux onto an absorbing linear chain. Considering the absorber to
be either a RW or a SAW the exponents read:
\begin{eqnarray} \label{5.5}
\lambda_{\mathrm{RW}}^{\mathrm{G}}(n) \def \lambda_{2,n}&=&-\eta^{\mathrm{G}}_{2n},
\\ \label{5.6}
 \lambda_{\mathrm{SAW}}^{\mathrm{U}}(n) \def
\lambda_{2,n}&=&-\eta^{\mathrm{U}}_{2n}+\eta_{20}.
\end{eqnarray}

Numerical values for the exponents $\lambda(n)$ may be easily
extracted from the numerical values of the exponents $\eta_{2n}$
given in tables \ref{tab2}, \ref{tab3}. Moreover, one can consider
an  absorber in the form of an $f_1$-arm star and generalize the
above considerations looking for the scaling laws of diffusing
particles in the vicinity of the core of such a polymer star.
Again, the scaling may be expressed in terms of the co-polymer star
exponents \cite{FerHol99}.

It is well known that polymer chains are fractals with a fractal
dimension defined by a correlation length exponent: $d_f=1/\nu$.
It appears that the behaviour of Brownian motion in the vicinity
of an absorbing star polymer possesses {\em multifractal} \cite{mf}
features. These are analyzed in detail in
\cite{FerHol99,FerHol97b}.

\section{Diffusion-controlled reactions in presence of polymers}\label{VI}

As mentioned in the preceding section, one of the  processes
described by equations (\ref{5.1}), (\ref{5.2}) may also be
considered as diffusion limited catalysis. In this section, we
will comment on a general phenomenon which can again be described
in terms of the star exponents: chemical reactions between
diffusing reactants. Examples of such reactions can be found in
different systems, ranging from biological systems to nuclear
reactors (see e.g. \cite {Rice85} and references therein). One
more place where these reactions appear as a limiting stage is the
aggregation models \cite{agr}. Of particular interest are the
reactions between reactants of different nature: particles $A$
which freely diffuse in a solution and particles $B$ which are
attached to polymer chains immersed in the same solvent
\cite{Oshanin} (the concentration of polymers being low enough to
allow to neglect the inter-chain interaction). Such a process may
be also considered as a trapping reaction of $n$ particles of $A$
type and  traps $B$:
\begin{equation}\label{6.1}
\label{1} A^n + B \rightarrow B.
\end{equation}
The reaction rate $k_n$ of (\ref{1}) in the vicinity of a certain
trap on the polymer of size $R$ is proportional to the averaged
moments of the concentration $\rho$ of diffusing particles near
this trap. From (\ref{5.3}) we get that the reaction rate scales
with $R$ as \cite{Cates}:
\begin{equation}\label{6.2}
\label{2} k_n \sim \langle\rho^n\rangle \sim (R/l)^{-\lambda_n},
\end{equation}
with $l$ being a characteristic length scale.

The scaling exponents $\lambda_n$ (\ref{6.2}) were considered in
the previous  section (see formulas (\ref{5.5}) and (\ref{5.6})).
Now, considering the absorber to be either a RW star or a SAW star
of $m$ chains, we define the exponents $\lambda_{mn}$ which can be
related to the familiar copolymer star exponents $\eta_{mn}$
(\ref{4.4}) via scaling laws:
\begin{eqnarray} \label{6.3}
\lambda^{\mathrm{RW}}_{mn} &=&-\eta^{\mathrm{G}}_{mn}, \nonumber\\
\lambda^{\mathrm{SAW}}_{mn}
&=& -\eta^{\mathrm{U}}_{mn}+\eta^{\mathrm{U}}_{m0}.
\end{eqnarray}
Let us note that the case $m=2$ corresponds to a trap located on
the chain polymer, whereas $m=1$ corresponds to a trap attached at
the polymer extremity.

Numerical values for the exponent $\lambda_{mn}$ as well as
perturbation theory expansions for them follow from tables
$\ref{tab2}$, $\ref{tab3}$ and are analyzed in detail in
\cite{FerHol97c,FerHol97,FerHol99a,FerHol01b}. Here, let us
analyze several particular cases for an absorbing $m$-star with a
reactant trap at its core and simply absorbing traps along all
chains:
\begin{itemize}
\item
{\em For a given $m$-star absorber} of size $R$, the reaction rate
(\ref{2}) scales as $k_{mn} \sim (R/l)^{-\lambda_{mn}}$. The increase
of the size $R$ by a factor of $a$ results in $k^{\prime}_{mn}
\sim (aR/l)^{-\lambda_{mn}}$ leading to:
\begin{equation}\label{6.4}
 k^{\prime}_{mn}/k_{mn} \sim a^{-\lambda_{mn}}
\end{equation}
with $\lambda_{mn}$ positive: increasing  $R$ by a factor of $a$
the reaction rate decreases $a^{-\lambda_{mn}}$ times due to the
increase of absorbing traps.
\item
{\em For a given reaction type } (\ref{1}) (i.e. for a fixed
number $n$ of particles which are trapped simultaneously)
attaching $m_1$ additional arms to an $m$-arm star absorber
results in a decrease of the reaction rate:
\begin{equation}\label{6.5}
 k_{(m+m_1),n}/k_{mn} \sim (R/l)^{-(\lambda_{(m+m_1),n}
-\lambda_{mn})},
\end{equation}
as far as $\lambda_{m_2n} > \lambda_{m_1n}$ for $m_2>m_1$.
\item
{\em For a given $m$-star absorber} the change of the type of
reaction (\ref{1}) to $A^{n_1} + B \rightarrow B, \, n_1>n$
results in a decrease of the reaction rate:
\begin{equation}\label{6.6}
 k_{m,(n+n_1)}/k_{mn} \sim (R/l)^{-(\lambda_{m,(n+n_1)}
-\lambda_{mn})},
\end{equation}
as far as $\lambda_{mn_2} > \lambda_{mn_1}$ for $n_2>n_1$; now,
more particles need to find the trap at the core simultaneously
for a reaction to take place.
\end{itemize}


\section{Conclusions and outlook} \label{VIII}

The notion of critical exponents is one of the central notions in
the theory of critical phenomena. The precise determination of the
set of critical exponents governing an equilibrium thermodynamic
second order phase transitions provides a challenge in numerous
studies. The exponents usually govern the power-law temperature
behaviour which manifests itself in the vicinity of the critical
point. In polymer theory, the direct ``relatives" of the exponents
governing temperature behaviour at the 2nd order phase transitions
are the exponents describing the configurational properties of
long flexible polymer chains. They govern the scaling laws
appearing in the limit of infinite chain length. These exponents
are important in defining the entropy of a polymer solution, the
radius of gyration of a polymer coil, the osmotic pressure of the
polymer component in a solution, its rheologic properties, etc
\cite{deGennes79,polymerRG}. Both
 the 2nd order phase transition critical exponents and the polymer chain
scaling exponents are closely related to each other and, finally,
correspond to the second derivative of the appropriate free energy.
However, in polymer theory there exist sets of exponents which do
not have direct ``relatives" in the theory of 2nd order phase
transitions. These exponents describe the scaling of star polymers
and polymer networks \cite{Duplantier89}. Whereas the scaling
exponents for chain polymers are calculated by now in different
techniques with the standard accuracy of modern theory of critical
phenomena, this is not the case for the star exponents.

In the present paper we have reviewed some recent calculations of
the critical exponents governing the scaling behaviour of star
polymers. To calculate the critical exponents it is today standard
to rely on the renormalization group technique improved by a
resummation of asymptotic series \cite{RGbooks,polymerRG}. The
numbers provided in tables~\ref{tab1}--\ref{tab4} of the paper
provide reliable data which may be used in different problems
where the star exponents become important. Some of the problems
were mentioned in the sections \ref{VII}--\ref{VI}:  the
interaction between star polymers in a good solvent, the Brownian
motion near absorbing polymers, diffusion-controlled reactions in the
presence of polymers.

\label{last@page}
\end{document}